\documentclass[runningheads]{llncs}
\usepackage{graphicx}
\usepackage{amssymb}
\usepackage{amsmath}
\usepackage{comment}
\usepackage[normalem]{ulem}
\usepackage[dvipsnames]{xcolor}
\usepackage{hyperref}

\begin{document}
%
%\title{Patches vs whole brain unsupervised Parkinson's disease anomaly detection
%\title{Patches vs whole brain approaches for subtle local unsupervised anomaly detection in MR scans. Application to early Parkinson Disease 
\title{Patch vs. global image-based unsupervised anomaly detection in MR brain scans of early Parkinsonian patients 
\thanks{
$\,$ Data  used  in  the  preparation  of  this  article  were  obtained  from  the  Parkinson’s  Progression  Markers  Initiative  (PPMI)  database  (www.ppmi-info.org/data).\\
· VMR is supported by a grant from NeuroCoG IDEX-UGA (ANR-15-IDEX-02). This work is partially supported by the French program “Investissement d’Avenir” run by the Agence Nationale pour la Recherche (ANR-11-INBS-0006).   \\
· VMR and NP contributed equally to this work.\\
}}
\titlerunning{Unsupervised Parkinson's disease anomaly detection}
\author{
Ver\'{o}nica Mu\~{n}oz-Ram\'{i}rez\inst{1,2}\orcidID{0000-0002-0975-7064} \and
Nicolas Pinon\inst{3}\orcidID{0000-0001-8817-9911} 
Florence Forbes \inst{2}\orcidID{0000-0003-3639-0226} \and
Carole Lartizen \inst{3}\orcidID{0000-0001-7594-4231}\and
Michel Dojat \inst{1}\orcidID{ 0000-0003-2747-6845} 
}

%index{Mu\~{n}oz-Ram\'{i}rez, Ver\'{o]nica}
%index{Pinon, Nicolas}
%index{Forbes, Florence}
%index{Lartizen, Carole}
%index{Dojat, Michel}

\authorrunning{V. Mu\~{n}oz-Ram\'{i}rez et al.}
\institute{
Univ. Grenoble Alpes, Inserm U1216, CHU Grenoble Alpes, Grenoble Institut des Neurosciences, 38000 Grenoble, France \\
\email{\{first.last\}@univ-grenoble-alpes.fr} \and
Univ. Grenoble Alpes, Inria, CNRS, Grenoble INP, LJK, 38000 Grenoble, France \\
\email{\{first.last\}@inria.fr} \and
Univ. Lyon, CNRS, Inserm, INSA Lyon, UCBL, CREATIS, UMR5220, U1206, F‐69621, Villeurbanne, France\\ 
\email{\{first.last\}@creatis.insa-lyon.fr}
}

\maketitle              % typeset the header of the contribution
\begin{abstract}
%The abstract should briefly summarize the contents of the paper in
%Although neural  networks have proven very successful in a number of image analysis tasks, their use in medical imaging remains difficult when targeting more subtle tasks than massive natural scenes classification or segmentation. 
Although neural networks have proven very successful in a number of medical image analysis applications, their use remains difficult when targeting subtle tasks such as the identification of barely visible brain lesions, especially given the lack of annotated datasets. Good candidate approaches are patch-based unsupervised pipelines which have both the advantage to increase the number of input data and to capture local and fine anomaly patterns distributed in the image, while potential inconveniences are the loss of global structural information. We illustrate this trade-off on Parkinson's disease (PD) anomaly detection comparing the performance of two anomaly detection models based on a spatial auto-encoder (AE) and an adaptation of a patch-fed siamese auto-encoder (SAE). On average, the SAE model performs better, showing that patches may indeed be advantageous.

\end{abstract}
\section{Introduction}
\label{sec:int}
Medical imaging represents the largest percentage of data produced in healthcare and thus a particular interest has emerged in deep learning (DL) methods to create support tools for radiologists to analyze multimodal medical images, segment lesions and detect subtle pathological changes that even an expert eye can miss. The vast majority of these methods are based on supervised models which require to be trained on large series of annotated data, time- and resource-consuming to generate.

Over the years, several publicly available neuroimaging databases have been curated and completed with annotations. Some of the most prominent ones are: MSSEG, for multiple sclerosis lesion segmentation \cite{msseg}; BRATS, for brain tumor segmentation \cite{brats}; ISLES, for ischemic stroke lesion segmentation \cite{isles}; and mTOP for mild traumatic injury outcome prediction \cite{mtop}.
Challenges, namely those of MICCAI, are organized regularly to showcase the latest technological advancements and push the community towards better performances. However, there are several neurological diseases seldom studied due to the small size and subtlety of the lesions they present. This is the case of vascular disease, epilepsy, and most neurodegenerative diseases in their early stages. The main challenge for such pathologies indeed, is to identify the variability of the pathological patterns on images where the lesion is barely seen or not visible.

Unsupervised methods are good candidates to tackle both the lack of annotated examples and the subtlety of brain scan anomalies \cite{baur_autoencoders_2020,Shinde2019}. They rely on networks that learn to encode normal brain patterns in such a manner that any atypical occurrence can be identified by the inability of the network to reproduce it. Auto-encoders (AE), variational auto-encoders (VAE) \cite{kingma_auto-encoding_2014} and generative adversarial networks (GAN) \cite{goodfellow_generative_2014}, have been extensively used as building blocks for unsupervised anomaly detection due to their ability to learn high-dimensional distributions \cite{baur_autoencoders_2020}.

Parkinson's Disease (PD) is a neurodegenerative disorder that is only identifiable through routine MR scans at an advanced stage. Nevertheless, the manifestation of non-motor symptoms, years before the apparition of the first motor disturbances,
suggests the presence of physio-pathological differences that could allow for earlier PD diagnosis.
PD afflicts patients for as many as one to two decades of their lives and current treatments can only attenuate some motor manifestations \cite{Zhao2010}. Therefore, reducing the gap between diagnosis and the onset of the neurodegenerative process is of paramount importance to identify personalized treatments that would significantly slow its natural progression. Unsupervised anomaly detection models are here employed to explore such challenging MR data analysis.

In a previous work \cite{Munoz_isbi}, we compared deterministic and variational, spatial and dense autoencoders for the detection of subtle anomalies in the diffusion parametric maps of \textit{de novo} (i.e., newly diagnosed and without dopaminergic treatment) PD patients from the PPMI database \cite{ppmi}. Our results, while preliminary, offered compelling evidence that DL models are useful to identify subtle anomalies in early PD, even when trained with a moderate number of images and only two parametric maps as input.

Our goal in this paper is to compare an improved anomaly detection pipeline based on a deterministic spatial auto-encoder, hereafter simply referred as AE, to an adaptation of patch-based siamese auto-encoder (SAE) proposed in \cite{Alaverdyan2020}. This architecture was originally intended to the detection of subtle epileptic lesions, application for which it achieved promising results.

One important difference between the two compared architectures is the dimension of the input and output data. While AE were trained on 2D transverse slices, thus capturing a global pattern in the image, SAE were trained on small patches sampled throughout the data, making them more suitable to capture fine patterns but losing global structural integrity. 
Through this comparison we aim to analyze the advantages of patch-fed architectures for the identification of subtle and local abnormalities as well as to propose an alternative for anomaly detection in moderate size image datasets. 

\section{Brain anomaly detection pipeline}
\label{sec:pipeline}
The anomaly detection task with auto-encoders can be formally posed as follows: 
\begin{itemize}
    \item An auto-encoder is first trained to reconstruct normal samples as accurately as possible. This network is composed of two parts: an encoder (1) that maps the input data into a lower dimensional latent space, assumed to contain important image features, and a decoder (2) that maps the code from the latent space into an output image. 
    \item When fed by an unseen image, this trained network produces a reconstructed image from its sampled latent distribution which is the counterpart 'normal' part of the input image. 
    \item \textit{Reconstruction error} maps, computed as the difference between the input and output images, are thus assumed to highlight anomalous regions of the input data.
    \item \textit{Anomaly scores} at the voxel, region of interest or image levels may then be derived from the post-processing of these \textit{reconstruction error} maps.
\end{itemize}

In this work, we present a general framework for unsupervised brain anomaly detection based on auto-encoders to produce reconstruction error maps and a novel post-processing step to derive per-region anomaly scores.

\subsection{Autoencoder architectures}

\begin{figure*}[t]
    \centering
    \includegraphics[width=0.9\textwidth]{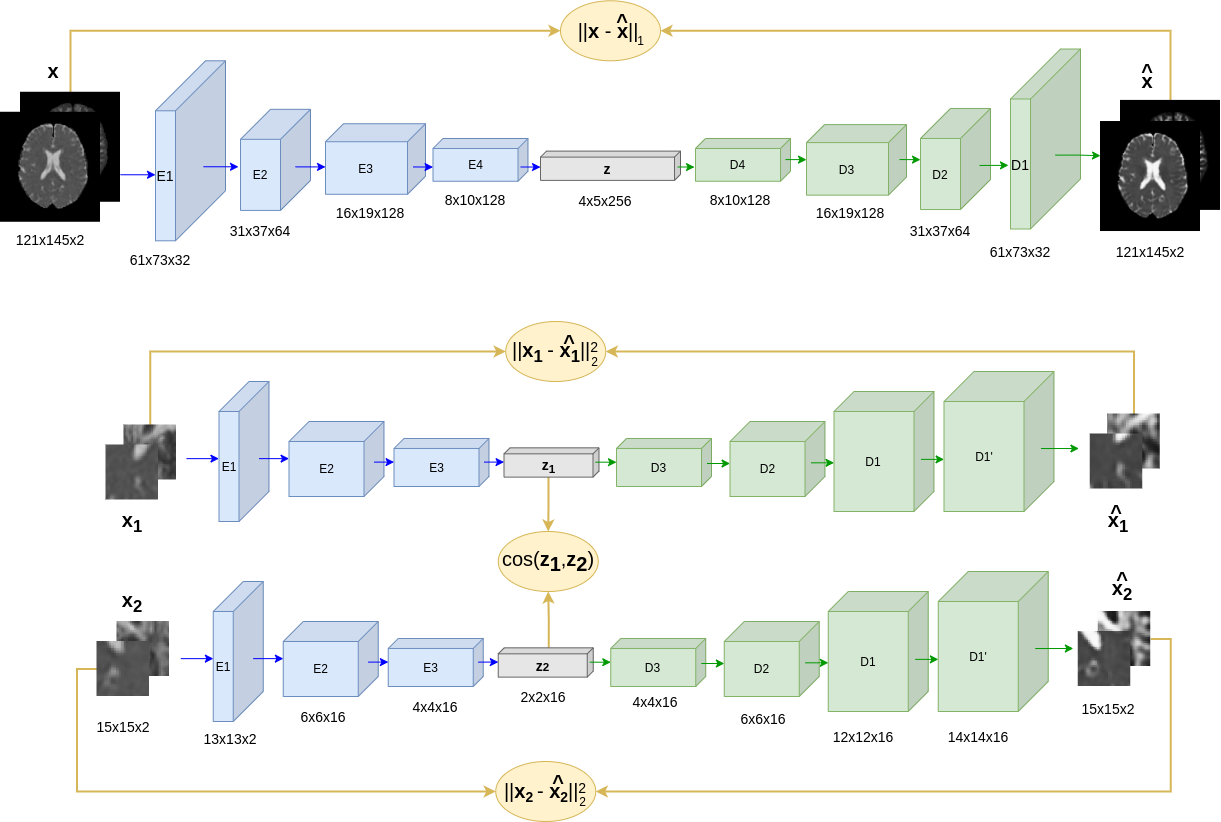}
    \caption{Classical auto-encoder (AE) on top, Siamese auto-encoder (SAE) at the bottom}
    \label{fig:archis}
\end{figure*}

We constructed and evaluated two auto-encoder models: a classic auto-encoder (AE) and a siamese auto-encoder (SAE). Both models are fully-convolutional. Their architectures are displayed in Figure \ref{fig:archis} and their differences are detailed below.

\subsubsection{Classic Auto-encoder:}
This architecture consists of 5 convolutional layers that go from input to bottleneck and 5 transposed convolutional layers going from bottleneck to output. The output of the encoder network is directly the latent vector $z$ and the loss function was simply the $L_1$-norm reconstruction error between input $\mathbf{x}$ and output $\hat{\mathbf{x}}$:
\begin{equation}
\label{eq:ae_loss}
L_{AE}(\mathbf{x}) = \|\mathbf{x} - \hat{\mathbf{x}}\|_1
\end{equation}

\subsubsection{Siamese Auto-encoder:} 
The siamese autoencoder (SAE) model \cite{Alaverdyan2020} 
consists of two identical convolutional autoencoders with shared parameters.
The SAE receives a pair of patches $(\mathbf{x_{1}}, \mathbf{x_{2}})$ at input that are propagated through the network, yielding representations $\mathbf{z_{t}} \in \mathcal{Z}, t = (1,2)$ in the middle layer bottleneck. The second term of the loss function $L_{SAE}$ (eq. \ref{eq:siamese_loss}) is designed to maximize the cosine similarity between $\mathbf{z_{1}}$ and $\mathbf{z_{2}}$. This constraints patches that are "similar" to be aligned in the latent space.
Unlike standard siamese architectures where \textit{similar} and \textit{dissimilar} pairs are presented to the network, Alaverdyan et al. \cite{Alaverdyan2020} proposed to train this architecture on \textit{similar} pairs only and compensate for the lack of \textit{dissimilar} pairs through a regularizing term that prevents driving the loss function to 0 by mapping all patches to a constant value. This term is defined as the mean squared error between the input patches and their reconstructed outputs. 
The proposed loss function for a single pair hence is:\\
\begin{equation}
\label{eq:siamese_loss}
L_{SAE}(\mathbf{x_1}, \mathbf{x_2}) =  \sum_{t=1}^{2} { {||\mathbf{x_{t}} -\mathbf{\hat{x}_{t}}||_2^2}} - \alpha \cdot  {cos(\mathbf{z_{1}}, \mathbf{z_{2}})}
\end{equation}
$ $\\
where $\mathbf{\hat{x}_{t}}$ is the reconstructed output of the patch $\mathbf{x_{t}}$ while $\mathbf{z_{t}}$ is its representation in the middle layer bottleneck and $\alpha$ an hyperparameter that controls the trade-off between the two terms.

As depicted on Figure \ref{fig:archis}, the encoder part is composed of 3 convolutional layers and one maxpooling layer in-between the first and second convolutions, while the (non symmetrical) decoder part is composed of 4 convolutional layers, with an upsampling layer in-between the second and third convolutional layer.

\subsection{Post-processing of the reconstruction error maps}

We leveraged the reconstruction error maps obtained from both architectures to generate an anomaly score, following the methodology introduced in \cite{Munoz_AIM}. The voxel-wise reconstruction errors in one image were computed as  $||x_i - \hat{x}_i||_1$. Since the architectures were fed more than one channel (here two MR modalities), we defined the joint reconstruction error of every voxel as the square root of the sum of squares of the difference between input and output for every channel. 

\begin{comment}
We thus defined the joint reconstruction error of every voxel as: 
\begin{equation}
\label{eq:joint_error}
e_i = \sqrt{(\text{FA}_i - \Hat{\text{FA}_i})^2 + (\text{MD}_i - \Hat{\text{MD}_i})^2}
\end{equation}
\end{comment}

Next, we fixed a threshold on these generated \textit{reconstruction error maps} to decide whether or not a given voxel should be considered as abnormal, hereafter called the \textit{abnormality threshold}. Since we expected PD patients to exhibit abnormal voxels in larger quantities than controls, this value corresponded to an extreme quantile (eg. the 98\% quantile) of the errors distribution in the control population.
The thresholded reconstruction error maps were then employed to identify anatomical brain regions for which the number of abnormal voxels could be used to discriminate between patients and controls.

\section{Experiments}
\label{sec:experiment}
\subsection{Data}

The dataset used in this work consisted of DTI MR scans of 57 healthy controls and 129 \textit{de novo} PD patients selected from the PPMI database. All images were acquired with the same MR scanner model (3T Siemens Trio Tim) and configured with the same acquisition parameters. Only one healthy control was taken out of the study due to important artifacts in the images.

From these images, two measures, mean diffusivity (MD) and fractional anisotropy (FA), were computed using MRtrix3.0.  Values of FA and MD were normalized into the range $[0, 1]$. The images were spatially normalized to the standard brain template of the Montreal Neurological Institute (MNI) with a non-linear deformation. The resulting MD and FA parameter maps were of dimension $121 \times 145 \times 121$ with a voxel size of $1.5 \times 1.5 \times 1.5 \text{mm}^3$. 

The control dataset was divided into 41 training controls and 15 testing controls to avoid data leakage. This division was effectuated in 10 different manners through a bootstrap procedure in order to assess the generalization of our predictions as advised in \cite{Poldrack2019}.
We took special care to maintain an age average around 61 years old for all the training and test population as well as a 40-60 proportion of females and males.

Once the models were trained with one of the 10 training sets, they were evaluated with the corresponding healthy control test set and the PD dataset (age: 62 y. $\pm$ 9; sex: 48 F).

\subsection{Training of the auto-encoders}

\subsubsection{AE models}
The training dataset of the AE models consisted of 1640 images corresponding to 40 axial slices around the center of the brain for each of the 41 training control subjects. 
The AE models were trained for 160 epochs, with a learning rate of  $10^{-3}$.  $3 \times 3$ kernels were convoluted using padding of 1 pixel and a stride of (2, 2). The bottleneck dimensions were h=4, w=5 and c=256. There were no pooling layers.
Implementation was done in Python 3.6.8, PyTorch 1.0.1, CUDA 10.0.130 and trained on a NVIDIA GeForce RTX 2080 Ti GPU with batches of 40 images. After each convolutional layer, batch normalization \cite{ioffe_batch_2015} was applied for its regularization properties. 
The rectified linear unit (ReLU) activation function was employed in each layer except the last, for which a sigmoïd was preferred. 
The loss functions were optimized using Adam \cite{kingma_adam_2017}. 

\subsubsection{SAE models}

The SAE model was trained with 600 000 patches of size  15$\times$15$\times$2 ($\sim$15 000 patches per subject).
The model was trained for 30 epochs, with a learning rate of $1\times 10^{-3}$. Bottleneck dimensions were h=2, w=2 and c=16. Maxpooling and upsampling layers were used, as detailled before. No batch-normalization layers were used. Activation function for every convolutional layer was the rectified linear unit (ReLU) and the sigmoid function was used in the last layer.
The kernel size and the numbers of filters were 3x3 and 16 respectively for all convolution blocks but the final one with 2x2 and 2 (equal to the number of channels) respectively.
The stride was 1 for all convolution blocks.
The maxpooling/upsampling factor was 2.
Implementation was done in Python 3.8.10, Tensorflow 2.4.1, 11.0.221 and trained on a NVIDIA GeForce GTX 1660 GPU with batches of 225 patches. The loss function was comprised of a reconstruction part (mean squared error) and a similarity measure (cosine similarity) weighted by a coefficient $\alpha$=0.005. The loss function was minimized using Adam \cite{kingma_adam_2017}.

\subsection{Performance evaluation}
The percentage of abnormal voxels found in the thresholded reconstruction error maps was employed to classify them as healthy or pathological (PD) based on a threshold. The critical choice of the threshold was investigated using a Receiver Operator Curve (ROC), 
taking into account the imbalanced nature of our test set (15 healthy and 129 PD).
Every point in the ROC corresponds to the sensitivity and specificity values obtained by a given threshold. As proposed in \cite{Munoz_AIM}, the choice of the optimal threshold, referred to as the \textit{pathological threshold}, was based on the optimal geometric mean, \textit{g-mean}=$\sqrt{Sensitivity \times Specificity}$.

Additionally, to help evaluate the localization of anomalies, two atlases were considered: the Neuromorphometrics atlas \cite{neuromorphometrics} and the MNI PD25 atlas \cite{Xiao2015}. The first was used to segment the brain into 8 macro-regions: subcortical structures, white matter and the 5 gray matter lobes (Frontal, Temporal, Parietal, Occipital, Cingulate/Insular). The latter was specifically  designed  for  PD  patients  exploration. It contains 8 regions: substantia nigra  (SN), red  nucleus (RN), subthalamic nucleus (STN), globus pallidus interna and externa (GPi, GPe), thalamus, putamen and caudate nucleus. 
For all of the before-mentioned regions of interest (ROI), we calculated the \textit{g-mean} for the associated pathological threshold, leading to the classification performance of our models. 

\section{Results}
\label{sec:results}
\begin{figure}[!ht]
    \centering
    \includegraphics[width=0.55\textwidth]{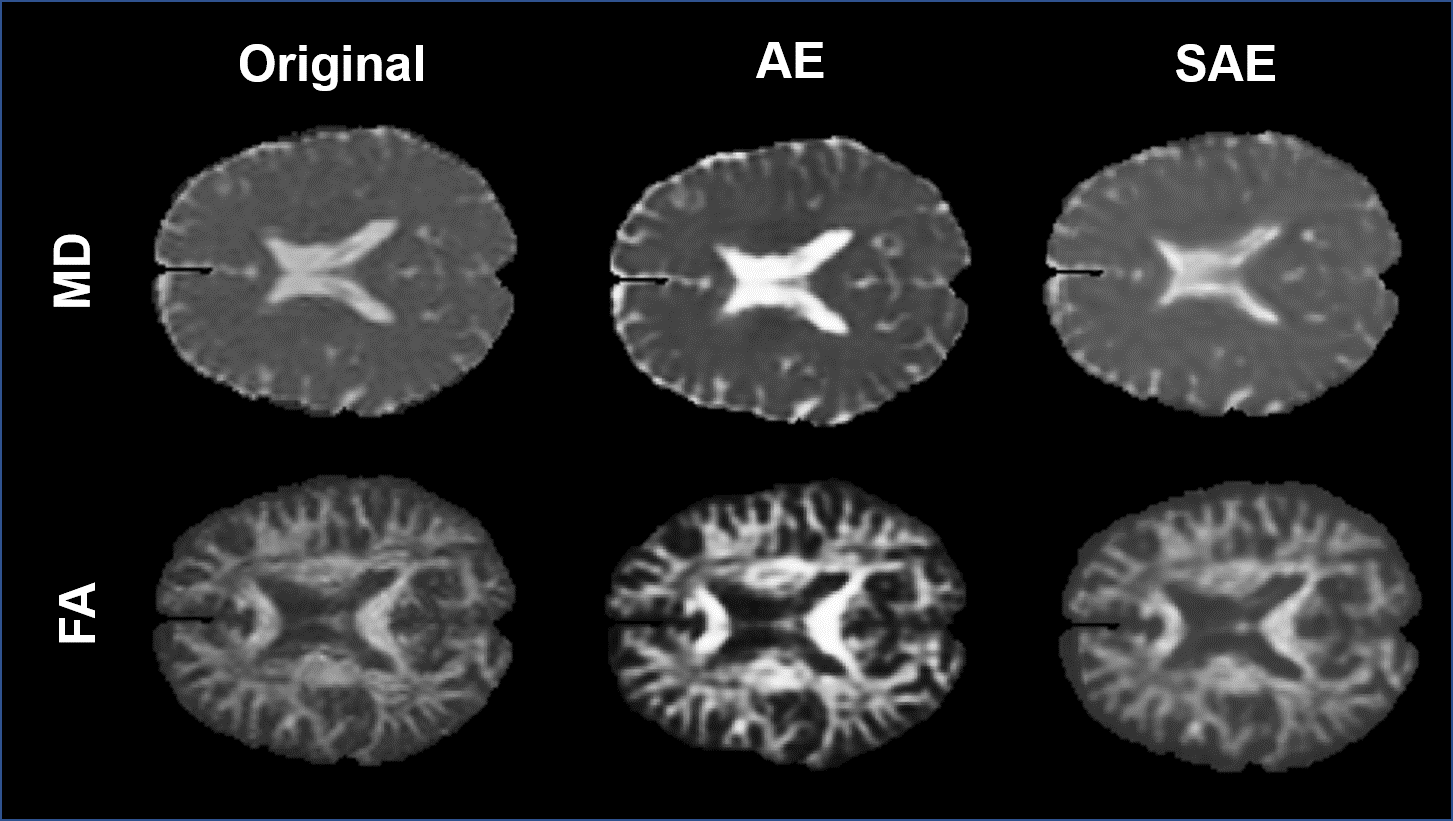}
    \caption{Showcase of a slice of the original data and its AE and SAE reconstructions}
    \label{fig:comp_recons}
\end{figure}

As it can be seen in Figure \ref{fig:comp_recons}, both auto-encoder architectures achieve good quality reconstructions, however the SAE seems to capture finer details and textures than the AE. This explains the high contrast in AE reconstructions.

\begin{figure}[!ht]
    \centering
    \includegraphics[width=\textwidth]{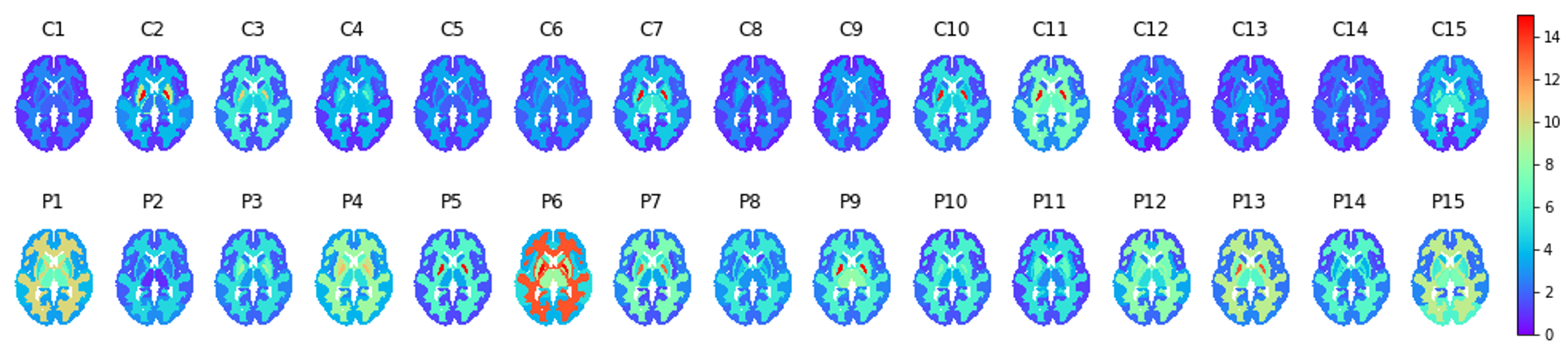}
    \caption{The percentage of abnormal voxels found by the SAE in the anatomical ROIs presented in Section \ref{sec:experiment}. Top: the test controls of Sample 1; Bottom: 15 randomly selected PD patients.}
    \label{fig:SAE_maps}
\end{figure}

The visualization of the percentage of abnormal voxels in the ROIs  presented in Figure \ref{fig:SAE_maps} showcases the inter-subject variability amongst members of the same population (healthy and PD). Even so, abnormal voxels are clearly more numerous in the PD patient population.

\begin{figure}[!ht]
    \centering
    \includegraphics[width=1\textwidth]{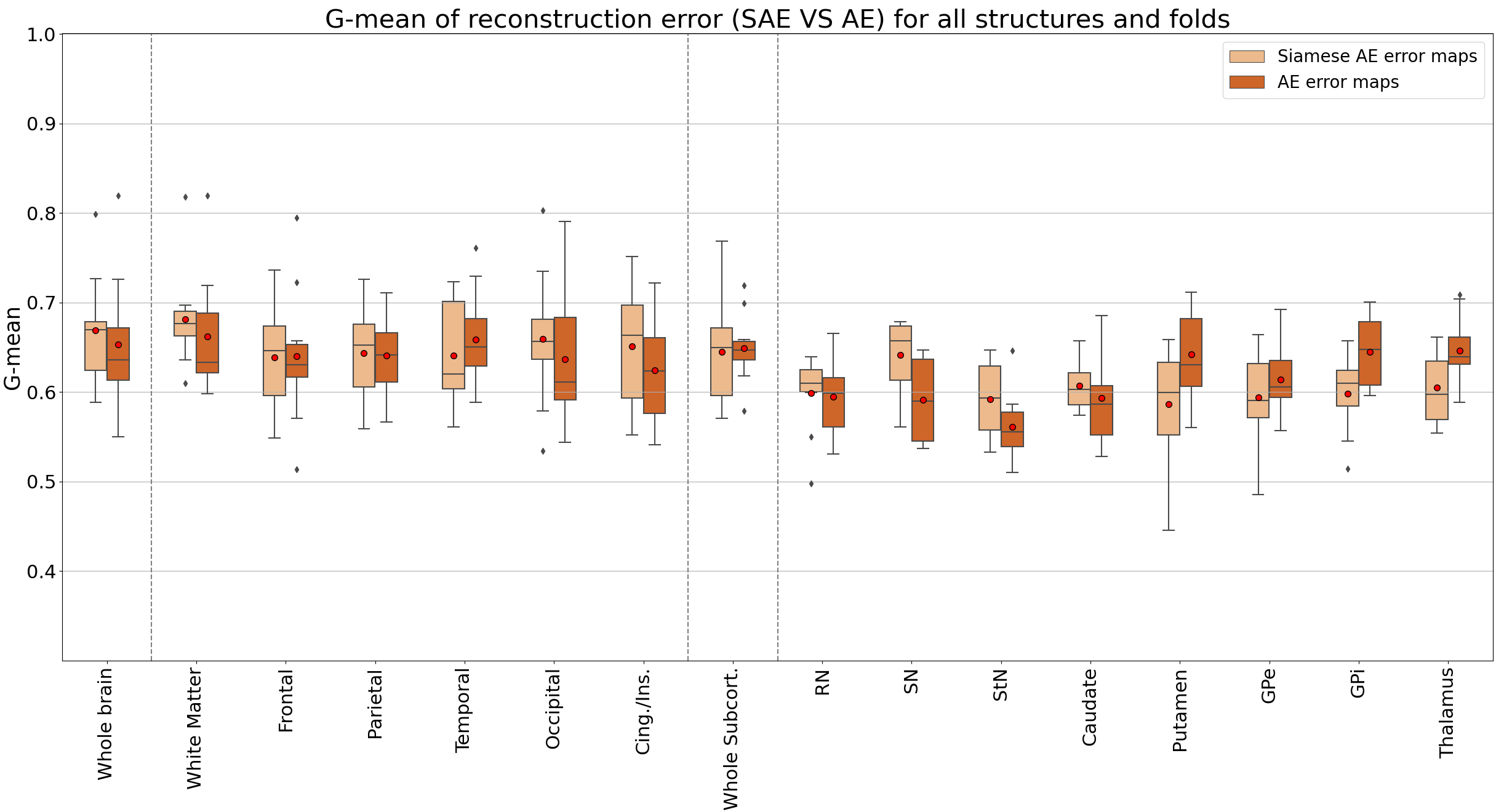}
    \caption{ \textit{g-mean} scores for the whole brain and several ROIs for AE and SAE. The vertical dashed lines separate macro- and micro brain structures. }
    \label{fig:gmean}
\end{figure}

The \textit{g-mean classification} scores for all models, obtained for each ROI and each sub-population, are presented in Figure \ref{fig:gmean}. We notice that, on average, the SAE model performed better than the AE on the whole brain and most of the macro and subcortical structures studied, with the exception of the temporal lobe, the putamen, the thalamus and the internal and external segment of the globus pallidus.
We note that the results varied greatly across the ten populations samples. As an indication, for the whole brain, the SAE obtained a \textit{g-mean} average score of $66.9 \pm 5.8\%$ and the AE $65.3 \pm 7.5\%$, however the best scores among the 10 samples were of $79.9\%$ and $81.9\%$ for the AE and SAE respectively, both on sample 1. Corresponding values for the white matter are of $68.2 \pm 4.6\%$ for SAE and $66.2 \pm 6.7\%$ for AE.
The largest standard deviations in the observed anatomical regions belonged to the white matter, the frontal and occipital lobes.

\section{Discussion and conclusion}
\label{sec:discu_conclu}
Unsupervised auto-encoders (AE) have shown to efficiently tackle challenging detection tasks where brain alteration are barely seen or not visible. The objective of our study was to explore the potential of such AE models for the detection of subtle anomalies in \textit{de novo} PD data and compare patch-based versus image-based models. 

Both AE and SAE architectures produced good quality reconstructions and were able to discriminate between healthy individuals and recently diagnosed PD patients with performances (see Figures 3 and \ref{fig:gmean}) that are competitive with those found in the literature. Notably, the Correia \& al. \cite{Correia2020} SVM mean accuracy score for a selection of WM regions is of 61.3\% whereas both the SAE and AE achieved g-mean scores above 66\% for the WM. Also, the cross-validation procedure of Schuff \& al. \cite{Schuff2015} obtained a ROC AUC of 59\% for the rostral segment of the SN which is below our SN average g-mean score for the SAE and equal to that of the AE.

Note that at this early stage of the disease (1-2 H-Y scale) the patients have no tremor nor uncontrolled movements compared to healthy subjects. This rules out that movement was the index that allowed PD classification. 

Using DTI data we did not search for structural atrophy or lesion load but for degradation of WM properties in the early stages of PD that could appear everywhere in the brain. This explains why the WM obtains the highest g-mean scores. This being said, our models could largely improve by increasing the size of our dataset. Furthermore, the addition of another MR modality such as iron load using T2/T2* relaxometry could allow us to detect the reduction of dopaminergic neurons in subcortical structures, largely reported in the early stages of PD but not visible in DTI.

Regarding the comparison between the AE and SAE, the choice is not clear. While the classic AE architecture benefits from a more straightforward implementation, the SAE proposes significant advantages for small databases. Indeed, patch-fed networks can be trained with smaller samples of data and the siamese constraint of the architecture ensures efficient learning. What is more, the latent space features of these models contain local information that can be used to classify between healthy and pathological individuals at the voxel-level and produce anomaly maps like those introduced by \cite{Alaverdyan2020}. 

In future work, we plan to generate this kind of maps for early PD patients to offer more precise indications about the localization of anomalies and correlate them with the PD hemispheric lateralization. In addition, we aim to complete our dataset by adding other MR modalities such as perfusion and relaxometry, but also by gathering heterogeneous data from multi-vendors and multi-sites exams. For this purpose, we will use an harmonization procedure as a  preprocessing step (extension of DeepHarmony \cite{dewey}). Finally, our 3D implementation of the SAE model is ongoing.

\bibliographystyle{splncs04}
\bibliography{refs}

\begin{comment}

\end{comment}

\end{document}